# Edge-Soliton-Mediated Vortex-Core Reversal Dynamics

Ki-Suk Lee, Myoung-Woo Yoo, Youn-Seok Choi, and Sang-Koog Kim[*]

*National Creative Research Center for Spin Dynamics & Spin-Wave Devices, Nanospinics Laboratory, Research Institute of Advanced Materials, Department of Materials Science and Engineering, Seoul National University,*

*Seoul 151-744, Republic of Korea*

We report a new reversal mechanism of magnetic vortex cores in nanodot elements driven by out-of-plane currents, occurring through two coupled edge-solitons via dynamic transformations between magnetic solitons of different topological charges. This mechanism differs completely from the well known switching process mediated by the creation and annihilation of vortex-antivortex pairs in terms of the associated topological solitons, energies, and spin-wave emissions. Strongly localized out-of-plane gyrotropic fields induced by the fast motion of the two coupled edge-solitons enable a magnetization dip that plays a crucial role in the formation of the reversed core magnetization. This work provides a new physical insight into the dynamic transformations of magnetic solitons in nanoelements.

Nontrivial inhomogeneous magnetization configurations in the restricted geometries of micrometer-size (or smaller) magnetic elements play crucial roles in the magnetization dynamics occurring on scales of a few tens of picoseconds [1-3]. In the past five years, such dynamics have attracted growing interest for their potential applications to future data storage [4-7] and processing devices [8]. For example, domain wall motions in the Walker breakdown regime can be described in terms of the sequential processes of the creation, propagation, and annihilation of various types of topological solitons, i.e., transverse, vortex and antivortex walls [3,9,10]. Also, a single magnetic vortex in nanodots can be transformed into itself but with a reversed core orientation (opposite to its original core orientation) via serial dynamic processes of the creation and subsequent annihilation of vortex-antivortex (VAV) pairs, as assisted by vortex-core gyration motions [11-18].

Furthermore, it has been reported that vortex-core reversals take place, avoiding vortex-core gyration motions, under a specific condition of an intended immobile vortex core [19] or by application of strongly localized pulse fields in the vicinity of the original core [20]. Regardless of the particular conditions and types of driving forces, the common reversal mechanism thus far found is the dynamic process of VAV pair creation followed immediately by the annihilation of the newly created antivortex and the original vortex inside the nanoelements. Such core reversals are always accompanied by an exchange energy explosion and subsequent spin-wave emissions phenomena.

In this Letter, we report a new vortex-core reversal mechanism driven by out-of-plane currents, which occurs through the creation of two coupled edge-solitons of half-integer winding number topological charge, not followed by exchange energy explosion and spin-wave emission phenomena. The physical origin of the edge-soliton-mediated vortex-core reversals are addressed.

Our approach in this study was to use micromagnetic numerical simulation, owing to its sufficient spatial and temporal resolutions and reliable resulting outputs for the given model's dimensional scale. We employed a model system constituting free-standing Permalloy (Py, $Ni_{80}Fe_{20}$) [21] thin-film disks of $2R = 200$ nm diameter and $L = 7$ nm thickness [Fig. 1(a)]. For vortex excitations up to its core switching, we used out-of-plane dc currents applied in the $+z$ direction (designated as $i_p = +1$) and a perpendicular polarizer of the spin-polarization direction $\hat{\mathbf{m}}_P = S_{pol}\hat{\mathbf{z}}$ with $S_{pol} = -1$. Such current flows produce two different effects: spin transfer torque (STT), and an in-plane circumferential Oersted field (OH) around the given current pass [22,23]. The OOMMF code (version 1.2a4) used [24] incorporates the Landau-Lifshitz-Gilbert equation [25] and an additional STT term: $\partial \mathbf{M}/\partial t = -\gamma\left(\mathbf{M}\times\mathbf{H}_{eff}\right) + \alpha/|\mathbf{M}|\left(\mathbf{M}\times\partial\mathbf{M}/\partial t\right) + \mathbf{T}_{STT}$ (Ref. [26]) with the phenomenological damping constant $\alpha$, the gyromagnetic ratio $\gamma$, and the effective field $\mathbf{H}_{eff}$. The STT is given by $\mathbf{T}_{STT} = (a_{STT}/|\mathbf{M}|)\mathbf{M}\times(\mathbf{M}\times\hat{\mathbf{m}}_P)$ with $a_{STT} = \frac{1}{2\pi}h\gamma P j_0/(\mu_0 2eM_s L)$, where $h$ is the Plank's constant, $j_0$ the current density, $\mu_0$ the vacuum permeability, $e$ the electron charge, $M_s$ the saturation magnetization, and $P = 0.7$ the degree of spin polarization. We assumed that in the current-perpendicular-to-plane geometry, currents flow uniformly through the entire Py disk; thus we calculated the circumferential OHs induced around the current pass simply using Ampere's law [22,23] [see Fig. 1(b)]. The initial ground state [27] of a magnetic vortex is the upward core magnetization, corresponding to polarization $p = +1$ and the counter-clockwise (CCW) in-plane curling magnetization, chirality $C = +1$. The other parameters applied were as follows: $\alpha = 0.01$, $\gamma = 2.21 \times 10^5$ m/As, and unit cell size $2 \times 2 \times 7$ nm$^3$.

Figure 2 shows the results of the numerical simulations. According to the current density $j_0$ of the out-of-plane dc currents, the observed dynamic behaviors were very distinct. We observed stationary vortex-core gyrations [23,29-31] with the CCW rotation sense for $1.15 \times 10^6$ A/cm$^2 \leq j_0 < 1.6 \times 10^6$, notably, persistently rotating motions of two *coupled edge-solitons* within the range of $1.6 \times 10^6$ A/cm$^2 \leq j_0 < 2.4 \times 10^6$ A/cm$^2$, and vortex-core reversals for $2.4 \times 10^6$ A/cm$^2 \leq j_0 < 1.8 \times 10^8$ A/cm$^2$. In the dynamic regime of the motion of the coupled edge solitons, hitherto unknown, the initial vortex core gyrates with a continuously increasing orbit radius, and when it reaches the disk boundary, it is then transformed into a quasi-uniform magnetization structure in a *C*-like deformation state. This transient process results in the two coupled edge-solitons that rotate constantly in the same CCW sense as that of the initial upward core, as evidenced by the corresponding solitons' trajectory for $j_0 = 2.0 \times 10^6$ A/cm$^2$ (Fig. 2, middle). In the range of $2.4 \times 10^6$ A/cm$^2 \leq j_0 < 1.5 \times 10^8$ A/cm$^2$, however, the persistent rotation motion of the edge-soliton pair observed in $1.6\times10^6$ A/cm$^2$ $\leq j_0 <$ $2.4\times10^6$ A/cm$^2$ is no longer stable, but this pair is transformed into the downward vortex core, $p = -1$ (Fig. 2, right). Once the reversed core ($p = -1$) is formed, it experiences a damped gyration with decreasing orbit radius in the CW rotation sense. Finally, the vortex-core reversal is completed through the creation and annihilation of the edge-soliton pair, rather than the well known VAV pair mediated mechanism reported in Refs. [11-18].

In order to elucidate the edge-soliton-mediated core reversal dynamics, we studied the details of the transformation subprocesses from the standpoint of topological solitons, as shown in Fig. 3. The edge-soliton-mediated mechanism consists of three distinct processes: 1) annihilation of a single vortex state once it reaches the disk boundary via core gyration, 2) formation of two coupled edge-solitons and 3) creation of a reversed core vortex through fusion of the coupled edge-solitons. The transition from process 1 to process 2 is spontaneous, as evidenced by no energy barrier to the annihilation of the initial vortex core when it reaches the edge boundary (see comparison of energy plots in Fig, 3). By contrast, the transition from process 2 to process 3 is not spontaneous, as evidenced by the emergence of the stationary motion of the coupled edge-solitons along the disk edge within a broad range of $1.6\times10^6$ A/cm$^2$ $\leq j_0 < 2.4\times10^6$ A/cm$^2$, and by the fact that the coupled edge-solitons in their persistent motion are transformed into the vortex of the reversed core only with further increase in $j_0$. This reveals that the edge-soliton pair has to overcome an exchange energy barrier for the formation of the reversed vortex core inside the disk. In this case, an exchange energy explosion and resultant spin-wave emissions as found in VAV mediated core reversals disappear and rather the demagnetization energy of the two edge solitons decreases when the reversed vortex core starts to form inside the disk.

From the two-dimensional (2D) topological perspective [32,33], the dynamic transformation from a single vortex to a coupled edge-soliton state (and vice-versa) occurs while conserving the total winding number ($n$) topological charge of the involved solitons [33]. A single vortex denoted as a bulk ($n = +1$) soliton decomposes into two coupled edge ($n = +1/2$) solitons, holding the total winding number, $n = +1$. However, another topological charge, the skyrmion number [34,35] defined as $q = np/2$ decays from the up core ($q = +1/2$) into two edge-solitons ($q = 0$). Subsequently, a reversed down core ($q = -1/2$) is created from the fusion of the two $q = 0$ edge solitons. In the well-known VAV-mediated reversals, the total +1 (or −1) skyrmion charge of the original upward (or downward) vortex core [$q = +1/2$ (or $-1/2$)] and the newly created downward (or upward) antivortex [$q = +1/2$ (or $-1/2$)] decays to 0 via their annihilation, accompanied by spin-wave emissions [14,35-37]. In contrast, in the edge-soliton mediated process the exchange energy of the $q = +1/2$ skyrmion, rather than drastically dissipating through spin-wave emissions as in VAV-mediated reversals, converts to the magnetostatic energy of the edge-solitons (see Fig. 3, plot of energies) in which process the magnetostatic energy is converted back to the exchange energy of the reversed vortex core. This is because the 0 skyrmion number of the edge soliton pair has a strong magnetostatic energy due to the magnetic free poles at the disk edge boundary.

Next, to examine the driving force behind the transformation of persistently rotating coupled edge-solitons into the reversed vortex core, we conducted further micromagnetic simulations [38] for various $j_0$ values. From the acquired numerical data, the spatial distribution of the local out-of-plane magnetization $m_z$, and the out-of-plane gyrotropic field $h_z$ induced by the kinetic energy of the coupled edge solitons, could be directly calculated [16], as shown in Fig. 4(a). The value of $h_z$ is expressed as $h_z = -1/M_s^2[\mathbf{M}\times(d\mathbf{M}/dt)]_z$, as derived from the time-derivative term of the LLG equation [39, 40]. Figure 4(a) provides snapshot plane-view images of both the local $h_z$ (right) and $m_z$ (left) averaged during the steady motion of the coupled edge-solitons, for three different values of $j_0$, for example. As can be seen, the $m_z$ and $h_z$ are strongly concentrated locally in between the two edge-solitons. The maximum $h_z$ value is found there, as shown by the red-color spot [Fig. 3(a), right, bottom]. Such concentrations grow higher with larger current densities, and correspondingly, the two coupled edge-solitons move closer, inducing largely concentrated in-plane curling magnetization distortion in between the two edge solitons, and then $h_z$. We therefore calculated the maxima of $h_z$ and $m_z$ as a function of $j_0$ up to vortex-core reversal. As seen in Fig. 4(b), as $j_0$ increases, the negative maxima of $h_z$ and $m_z$ grow with the velocity of the coupled edge-soliton motion. In this case, the velocity of the coupled edge-solitons' motion reaches 531 ± 153 m/s, higher than the critical velocity (about 330 ± 37 m/s for Py) required for VAV-meditated core reversals [17,41]. With increases of $j_0$ beyond its critical value (here $j_{0,cri} = 2.3 \times 10^6$ A/cm$^2$), the strength of $h_z$ reaches its critical value, $h_{z,cri} = -2.1 \pm 0.2$ kOe, which value is required for a locally concentrated $m_z$ to become a magnetization dip in the direction opposite to the original core magnetization for the formation of the reversed core orientation as reported in our earlier works [16,40]. Here, the STT effect and in-plane OHs do not effect directly the out-of-plane magnetization distortion around the coupled edges. Only, sufficiently large in-plane OHs induce the in-plane curling magnetization distortion, which in turn makes $h_z$.

To conclude, we discovered, for the first time, a new reversal mechanism of magnetic vortex cores in nanodisks driven by out-of-plane currents. This reversal mechanism occurs through dynamic transformations between two coupled half-integer edge-solitons and integer bulk vortex solitons. The mechanism differs completely from the well known VAV reversal process in the types of the associated topological solitons, energies, and spin-wave emissions. This work provides a fundamental understanding of the dynamic transformations of magnetic topological solitons in nanoelements.

**Acknowledgements**

This work was supported by the Basic Science Research Program through the National Research Foundation of Korea (NRF) funded by the Ministry of Education, Science and Technology (grant no. 20100000706).

**References**

* To whom all correspondence should be addressed: sangkoog@snu.ac.kr

[1] A. P. Malozemoff and J. C. Slonczewski, Magnetic Domain Walls in Bubble Material (Academic, New York, 1979); V. G. Bar'yahhtar *et al.*, Dynamics of Topological Magnetic Solitons (Springer-Verlag, Berlin, 1994); A. Hubert, and R. Schäfer, *Magnetic Domains* (Springer-Verlag, Berlin, New York, Heidelberg, 1998).

[2] L. Thomas *et al.*, Nature **443**, 197 (2006)

[3] J.-Y. Lee *et al.*, Phys. Rev. B **76**, 184408 (2007).

[4] S. S. P. Parkin, M. Hayashi, L. Thomas, Science **320**, 190 (2008).

[5] R. P. Cowburn, Nature Mater. **6**, 255 (2007); J. Thomas, Nature Nanotech. **2**, 206 (2007).

[6] S.-K. Kim, K.-S. Lee, Y.-S. Yu, and Y.-S. Choi, Appl. Phys. Lett. **92**, 022509 (2008); S.-K. Kim, K.-S. Lee, Y.-S. Choi, and Y.-S. Yu, IEEE Trans. Magn. **44**, 3071 (2008)

[7] S. Bohlens, B. Kruger, A. Drews, M. Bolte, G. Meier, and D. Pfannkuche, Appl. Phys. Lett. **93**, 142508 (2008).

[8] D. A. Allwood *et al.*, Science **309**, 1688 (2005).

[9] K. Y. Guslienko, J.-Y. Lee, and S.-K. Kim, IEEE Trans. Mag. **44**, 3079 (2008)

[10] D. J. Clarke *et al.*, Phys. Rev. B **78**, 134412 (2008).

[11] B. Van Waeyenberge *et al.*, Nature **444**, 461 (2006).

[12] R. Hertel *et al.*, Phys. Rev. Lett. **98**, 117201 (2007). Q. F. Xiao *et al.*, J. Appl. Phys. **102**,

103904 (2007).

[13] K.-S Lee, K. Y. Guslienko, J.-Y. Lee, and S.-K. Kim, Phys. Rev. B **76**, 174410 (2007).

[14] S. Choi *et al.*, Phys. Rev. Lett. 98, 087205 (2007).

[15] K. Yamada *et al.*, Nature Mater. **6**, 269 (2007).

[16] K. Y. Guslienko, K.-S. Lee, and S.-K. Kim, Phys. Rev. Lett. **100**, 027203 (2008).

[17] K.-S. Lee *et al.*, Phys. Rev. Lett. **101**, 267206 (2008).

[18] S.-K. Kim *et al.*, Appl. Phys. Lett. **91**, 082506 (2007).

[19] V. P. Kravchuk, *et al.* Phys. Rev. B **80**, 100405 (2009).

[20] S. Gliga *et al*. 11th Joint Magnetism and Magnetic Materials-Intermag Conference, Washington DC, 2010, Oral GD-07 (unpublished).

[21] The material parameters for Py in this simulation were the magnetization saturation $M_s = 8.6 \times 10^5$ A/m, the exchange stiffness $A_{ex} = 1.3 \times 10^{-11}$ J/m, and a zero magnetocrystalline anisotropy constant.

[22] Y.-S. Choi, S.-K. Kim, K.-S. Lee, and Y.-S. Yu, Appl. Phys. Lett. **93**, 182508 (2008).

[23] Y.-S. Choi, K.-S. Lee, and S.-K. Kim, Phys. Rev. B **79**, 184424 (2009).

[24] See http://math.nist.gov/oommf.

[25] L. D. Landau and E. M. Lifshitz, Phys. Z. Sowjet. **8**, 153 (1935); T. L. Gilbert, Phys. Rev. **100**, 1243 (1955).

[26] J. C. Slonczewski, J. Magn. Magn. Mater. **159**, L1 (1996); L. Berger, Phys. Rev. B **54**, 9353

(1996).

[27] To excite the vortex gyration mode by out-of-plane dc currents, the initial vortex core was displaced 2 nm in the –$x$ direction from the dot center.

[28] Y.-S. Choi *et al.*, Appl. Phys. Lett. **96**, 072507 (2010).

[29] B. A. Ivanov and C. E. Zaspel Phys. Rev. Lett. **99**, 247208 (2007)

[30] A. V. Khvalkovskiy *et al.* Phys. Rev. B **80**, 140401 (2009).

[31] A. Dussaus *et al.*, Nature Comm. **1**: 8 doi: 10.1038/ ncomms1006 (2010).

[32] N. D. Mermin, Rev. Mod. Phys. **51**, 591 (1979).

[33] O. Tchernyshyov and G.W. Chern, Phys. Rev. Lett. **95**, 197204 (2005).

[34] T.H.R. Skyrme, Proc. R. Soc. London **262**, 237 (1961).

[35] O. A. Tretiakov and O. Tchernyshyov, Phys. Rev. B **75**, 012408 (2007).

[36] K.-S. Lee, S. Choi, and S.-K. Kim, Appl. Phys. Lett. **87**, 192502 (2005).

[37] R. Hertel and C. M. Schneider, Phys. Rev. Lett. **97**, 177202 (2006).

[38] To reduce computational time, we intended to form two edge-solitons by application of a 400 Oe static external field in the +$y$ direction. This field would allow for *C*-state magnetization deformation.

[39] A. A. Thiele, Phys. Rev. Lett. **30**, 230 (1973).

[40] M.-Y Yoo *et al.*, Phys. Rev. B (inpress).

[41] A. Vansteenkiste *et al.*, Nature Phys. **5**, 332 (2009).

**Figure captions**

FIG. 1. (Color online) Schematic illustration of model system. (a) Free-standing Py disk of indicated dimensions. (b) Spatial distribution of strength of OHs induced by current flow of density $j_0 = 2.0 \times 10^6$ A/cm$^2$. The in-plane rotation sense is indicated by the wide arrow, here CCW for $i_p = +1$.

FIG. 2. (Color online) Representative distinct vortex excitations driven by out-of-plane dc currents in different regions of $j_0$. For each region we chose $j_0 = 1.5$, 2.0, and 3.0, respectively. Top) Snapshot perspective images of temporal evolution of vortex dynamics for each $j_0$. The height of the surface indicates the out-of-plane components of the local magnetizations, and the streamlines with arrows indicate the in-plane components. Bottom) Trajectories of in-plane motions of upward (red) vortex core and its reversed downward (blue) core, or coupled edge-solitons

FIG. 3. (Color online) Comparison of two different dynamic behaviors of two coupled edge-solitons: (a) in-plane persistent rotation of two edge-solitons at disk boundary and (b) transformation to vortex core of reversed orientation. The color coding indicates the out-of-plane components of the local magnetizations, while the streamlines with arrows indicate the in-plane components. The dotted red arrows represent the directions of the motion of the coupled

edge-solitons. The right column shows the exchange ($E_{ex}$), demagnetization ($E_{de}$) energies and their sum ($E_{tot}$), and velocity variations, where the vertical lines indicate corresponding snapshot images. The orange horizontal lines represent the value $v = 350$ m/s.

FIG. 4. (Color online) (a) Spatial distributions of magnetization deformation, $m_z$ (left) and normal component of associated out-of-plane gyrotropic field, $h_z$ (right), for indicated $j_0$ values, as indicated by color codes. The streamlines with arrows indicate the in-plane local magnetizations. (b) Plots of minima of magnetization dip $m_{z,dip} = M_{z,dip}/M_s$ and local $h_z$ as functions of $j_0$ along with the velocity of motion of edge-solitons and associated energies. The vertical dotted line indicates the boundary above which value vortex-core reversals occur through the two coupled edge-solitons. Above the dotted vertical line, each value for vortex was estimated at the time when the reversed vortex core starts to gyrate with damping.

**FIG. 1.**

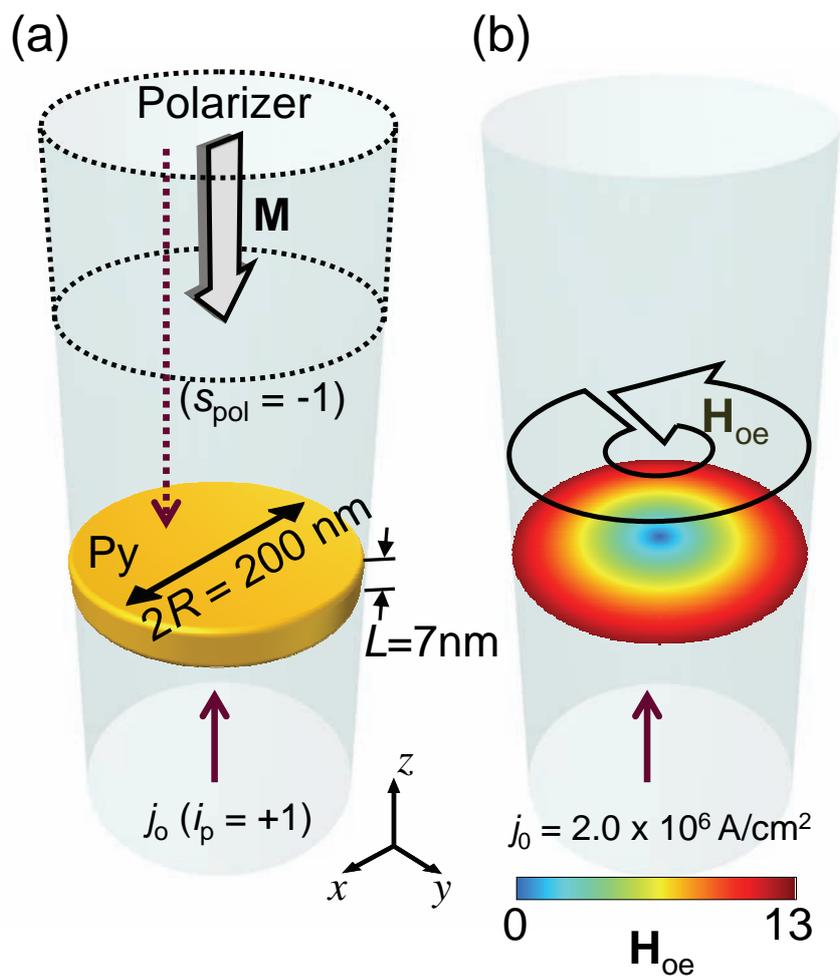

**FIG. 2.**

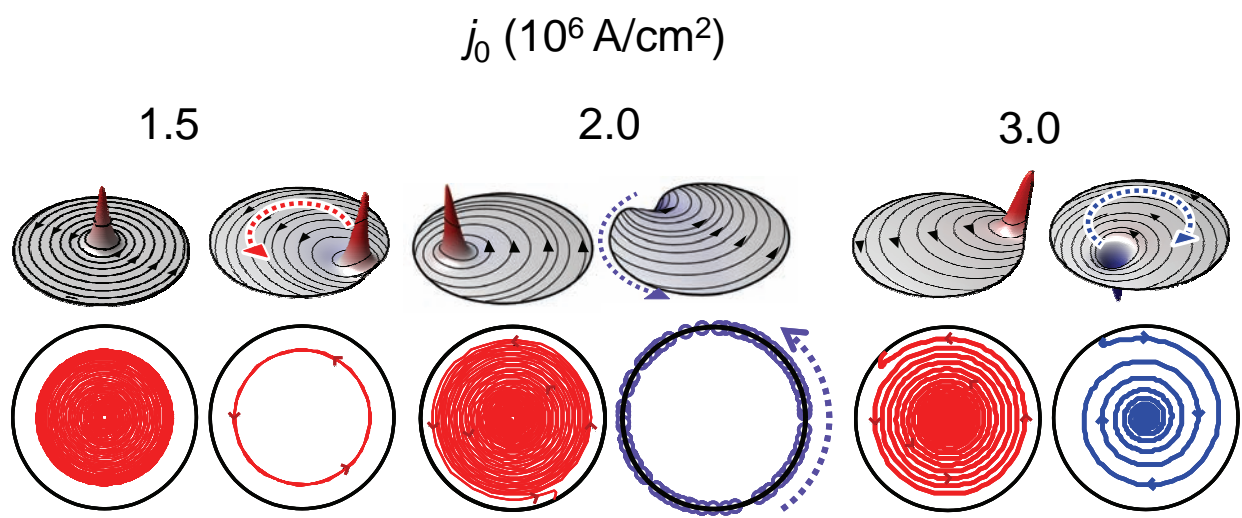

**FIG. 3.**

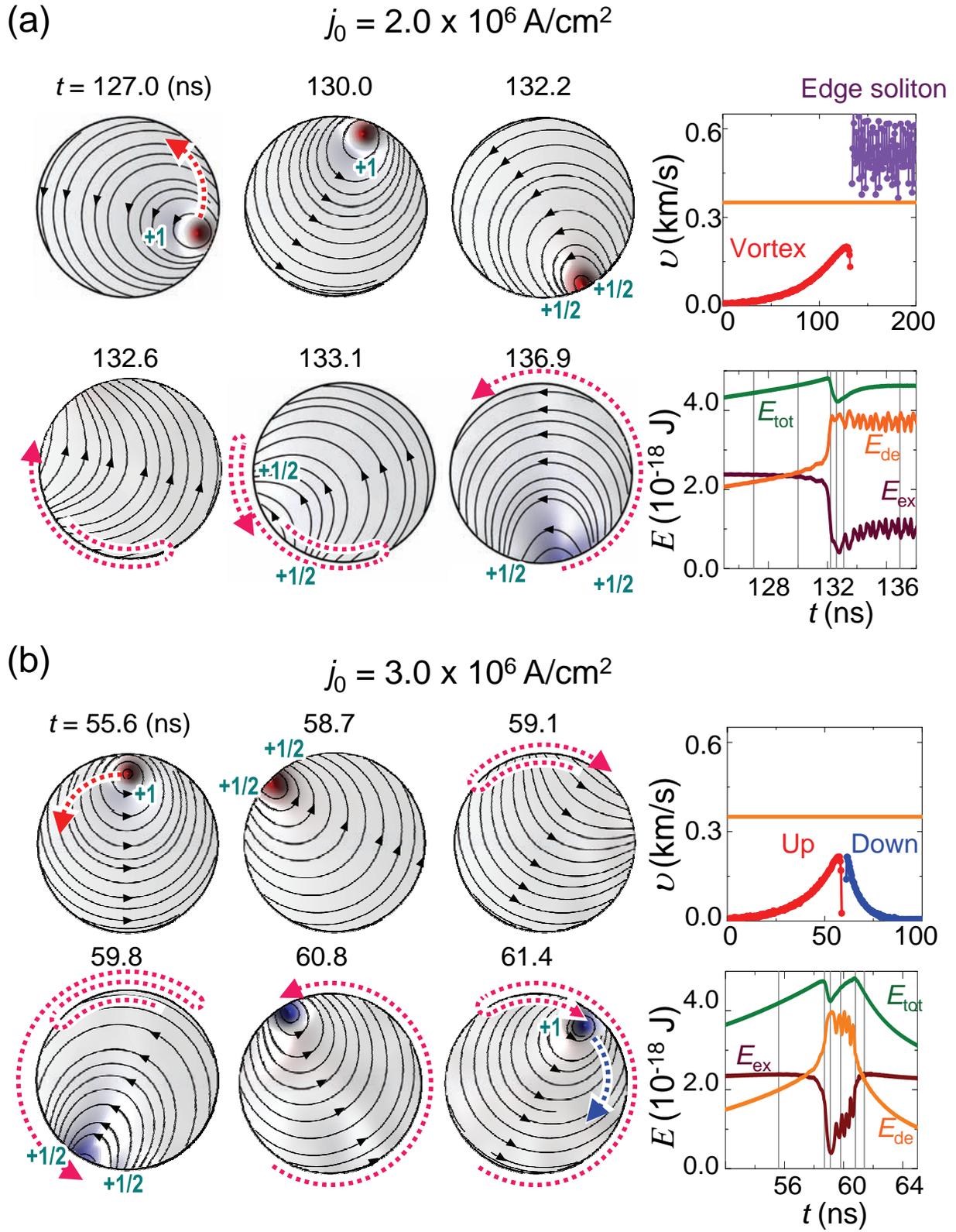

**FIG. 4.**

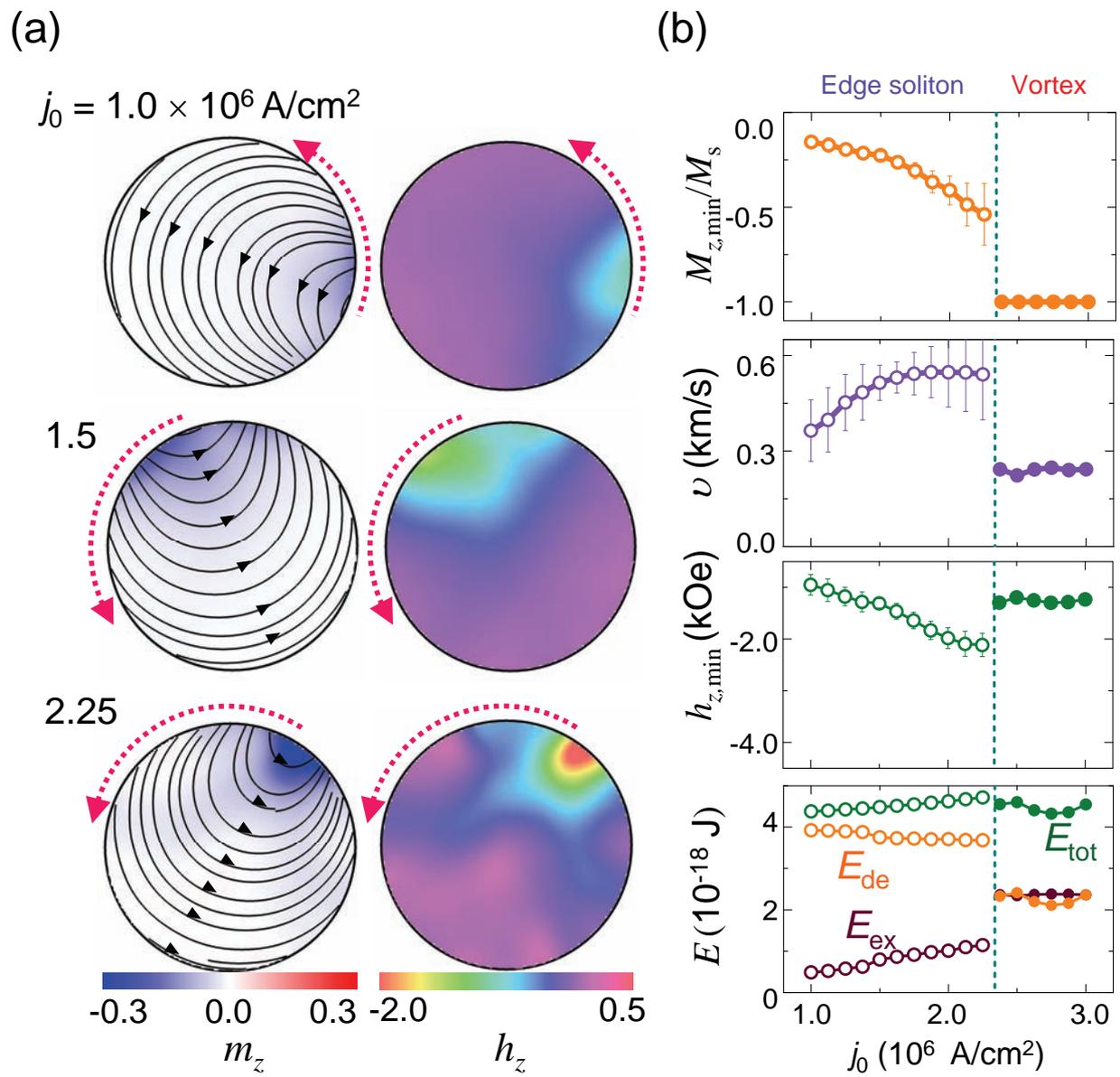